\documentclass[traditabstract]{aa}
\usepackage{graphicx}
\usepackage{txfonts}

\begin{document}

\title{Turbulent Motions and Shocks Waves in Galaxy Clusters simulated with AMR}
\author{F. Vazza\inst{1,2}, G. Brunetti\inst{2}, A. Kritsuk\inst{3}, R. Wagner\inst{3,4}, C. Gheller\inst{5}, M .Norman\inst{3,4}}

\offprints{Franco Vazza \\ \email{vazza@ira.inaf.it}}

\institute{Dipartimento di Astronomia, Universit\'a di Bologna, via Ranzani
1,I-40127 Bologna, Italy
\and  INAF/Istituto di Radioastronomia, via Gobetti 101, I-40129 Bologna,
Italy
\and Ctr. for Astrophysics and Space Sciences,
  U.C.~San Diego, La Jolla, CA 92093 
\and  Physics Department,
  U.C.~San Diego, La Jolla, CA 92093
\and CINECA, High Performance System Division, Casalecchio di
Reno--Bologna, Italy}

\date{Received / Accepted}

\authorrunning{F. Vazza et al.}
\titlerunning{Turbulence and Shocks in Galaxy Clusters}

\abstract
{We have implemented an Adaptive Mesh Refinement criterion explicitly designed
to increase spatial resolution around discontinuities in the velocity field in ENZO cosmological simulations.
With this technique, shocks and turbulent eddies developed during the
hierarchical assembly of galaxy clusters are followed with unprecedented
spatial resolution, even at large distances from the clusters center.
By measuring the spectral properties of the gas velocity field, its time evolution
and the properties of shocks for a reference 
galaxy cluster, we investigate the  connection between accretion processes
and the onset of chaotic motions in the simulated Inter Galactic
Medium over a wide range of scales.}

\maketitle

\keywords{galaxies: clusters, general -- methods: numerical -- intergalactic medium -- large-scale structure of Universe}

\section{Introduction}
\label{sec:intro}

The intergalactic medium (IGM) in galaxy clusters is likely turbulent, at
some level: this is claimed from several independent theoretical and numerical
approaches (e.g. Bryan \& Norman 1998; Ricker \& Sarazin 2001; Brunetti et al.2001; Inogamov \& Sunyaev 2003; Dolag et al.2005;
Subramanian et al.2006; Vazza et al.2006; Brunetti \& Lazarian 2007;
Nagai et al.2007; Iapichino \& Niemeyer 2008).
A number of observational evidences has also been published 
in the last few years. Using a mosaic of
XMM-Newton observations of the Coma cluster,
Schuecker et al.(2004) obtained spatially-resolved gas
pressure maps which indicate the presence of a
significant amount of turbulence, with a spectrum of the fluctuations 
consistent with a Kolmogorov turbulence.
Additional evidences of turbulent motions inside 
nearby galaxy clusters came from the observation of pseudo-pressure fluctuations in Abell~754 using XMM (Henry et al.2004) and from the non detection
of resonant scattering in the Perseus cluster (Churazov et al.2004).
Also studies of Faraday Rotation allow a complementary approach and suggest
that the IGM magnetic field is turbulent on a broad range of scales
(Murgia et al.2004; Govoni et al.2006; Ensslin \& Vogt (2006).

Detailed X-ray analysis performed in nearby cool-core galaxy clusters 
(e.g. Fabian et al.2003; Churazov et al.2004; Graham et al.2006; Ota et al.
2006) suggest that the turbulent velocity field is subsonic at the
scale of the cluster cores. 
Also, limits to the amount of turbulence in the IGM were recently
derived by  Churazov et al.(2008), suggesting that the amount of 
non-thermal pressure within $\sim 50$ kpc from the central galaxies
in Perseus and Virgo clusters cannot exceed $\sim 10-20$ per cent 
of the thermal energy budget at the same radius.

In addition, the phenomenology of diffuse radio halo emission suggests
a scenario in which turbulent MHD modes, excited during cluster mergers, 
may re-accelerate the relativistic emitting particles
(e.g. Ferrari et al.2008; Brunetti et al.2008; Cassano 2009 and
references therein).
Remarkably, the interplay between Cosmic Rays (CR) and turbulent magnetic
fields may drive still poorly explored plasma processes that 
may potentially affect our simplified view of the IGM (Subramanian et al.
2006, Schekochihin et al.2007, Brunetti \& Lazarian 2007, Guo \& Oh 2008).
From the theoretical point of view,
turbulence can be injected in the IGM by several mechanisms: 
plasma instabilities, cluster mergers and
shock waves, wakes of galaxies moving into the IGM, outflows from AGNs hosted in the center of galaxy clusters and galactic
winds.
The total energy budget in form of turbulent motions inside galaxy clusters, as well
as their distribution and their connection with cluster dynamics and non
gravitational process in galaxy clusters are sill open fields and cosmological
numerical simulations are potentially able to provide a great insight
in the characterization of the above phenomena.

Early Eulerian numerical simulations of merging clusters (e.g.,
Bryan \& Norman 1998; Ricker \& Sarazin 2001)
provided the first reliable representations of the way in which 
turbulence is injected into the IGM by merger events. More recently, high resolution
Lagrangian (Dolag et al.2005, Vazza et
al.2006) and Eulerian simulations (Nagai et al.2007, Iapichino
 \& Niemeyer 2008) found that a sizable
amount of pressure support (i.e. $\sim 10$ percent of the total pressure inside $0.5 R_{vir}$) 
in the IGM
is sustained by chaotic motions. 
Also, the amount of turbulent energy
delivered by mergers and accretions is found to 
scale with the thermal energy of simulated galaxy
clusters (Vazza et al.2006).

Despite the tremendous capability that Lagrangian simulations
have in resolving the smallest structures within galaxy clusters, they 
may suffer of serious limitations in
modeling fluid instabilities, mostly because of the
effects played by the artificial viscosity employed
to solve hydro equations (e.g. Agertz et al.2007; Tasker
et al.2008; Mitchell et al.2009). Therefore, the use of an 
Eulerian scheme free of artificial
viscosity, as the Piecewise Parabolic Method adopted in the ENZO, can provide an important
insight in all the above points. On the other hand Eulerian schemes 
with fixed grid resolution 
are typically limited by their low spatial resolution so that the application
of Adaptive Mesh Refinement (AMR) techniques is mandatory to achieve
adequate spatial detail in the simulations.

In this work we present first results from the application of an
Adaptive Mesh Refinement (AMR) to ENZO simulations, 
which allows to follow at the same time 
shocks and turbulent motions with unprecedented
resolution up to large distances from the galaxy cluster
center.
For the simulations presented here, we assume a $\Lambda$CDM cosmology with
parameters $\Omega_0 = 1.0$, $\Omega_{BM} = 0.0441$, $\Omega_{DM} =
0.2139$, $\Omega_{\Lambda} = 0.742$, Hubble parameter $h = 0.72$ and
a normalization of $\sigma_{8} = 0.8$ for the primordial density power
spectrum.

\begin{table}
\label{tab:tab1}
\caption{Main characteristics of the runs. ''D'' stands for AMR  
triggered by gas/DM 
over-density, while ''V'' stands for AMR triggered by velocity jumps. $\Delta$ is the peak gas spatial resolution. $\delta$ 
specifies the value adopted to trigger AMR, see Sec.3 for explanations.}
\centering \tabcolsep 5pt 
\begin{tabular}{c|c|c|c|c}
 ID & $N_{grid}$ &  $M_{dm}$ [$M_{\odot}/h$] & $\Delta$ [kpc] & AMR \\ \hline
 v256-4 & $256^{3}$ & $6.76 \cdot 10^{8}$ & $18$ & D+V($\delta>10$) \\ 
 v256-3 & $256^{3}$ & $6.76 \cdot 10^{8}$ & $36$ & D+V($\delta>3$) \\
 v128-3 & $128^{3}$ & $5.39 \cdot 10^{9}$& $36$ & D+V($\delta>3$) \\
 v64-3 & $64^{3}$ & $4.32 \cdot 10^{10}$& $36$ & D+V($\delta>3$) \\
 d128 & $128^{3}$ & $5.39 \cdot 10^{9}$& $36$ & D \\
 v128-10 & $128^{3}$ & $5.39 \cdot 10^{9}$ & $36$  & D+V($\delta>10$) \\
 v128-1 & $128^{3}$ & $5.39 \cdot 10^{9}$& $36$  & D+V($\delta>1$) \\
 v128-z2 & $128^{3}$ & $5.39 \cdot 10^{9}$& $36$  & D+V($\delta>3$,$z>2$) \\
 
\end{tabular}
\end{table}

\begin{figure} 
\includegraphics[width=0.49\textwidth]{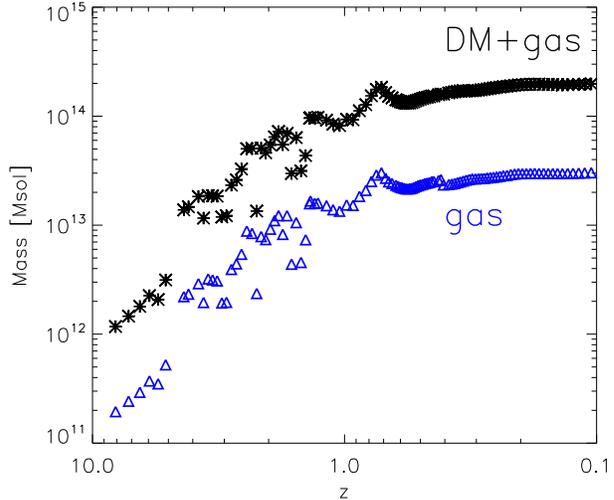}
\caption{Redshift evolution of the Dark matter plus gas mass ({\it black}), and of total gas mass ({\it blue}) inside the virial radius of the galaxy cluster studied in this work.}
\label{fig:mass_z}
\end{figure}

\section{Numerical Code and Setup}
\label{sec:simulations}

ENZO is an adaptive mesh refinement (AMR) cosmological hybrid 
code highly optimized for supercomputing
(Bryan \& Norman 1997, 1998; Norman \& Bryan 1999; Bryan, Abel, \& 
Norman 2001, O'Shea et al.2004; Norman et al.2007). It couples
an N-body particle-mesh solver with an adaptive mesh method for ideal 
fluid-dynamics (Berger \& Colella, 1989).
ENZO adopts an Eulerian hydrodynamical solver
based on the the Piecewise Parabolic 
Method (PPM, Woodward \& Colella, 1984),
that is a higher order extension of Godunov's shock capturing
method (Godunov 1959).
The PPM algorithm belongs to a class
of schemes in which an accurate representation of flow discontinuities is made
possible by building into the numerical method the calculation of the
propagation and interaction of non--linear waves. 
It is at least second--order accurate in space (up
to the fourth--order, in the case of smooth flows and small time-steps) and
second--order accurate in time. 
The PPM method describes shocks with high accuracy 
and has no need of artificial viscosity, leading to an 
optimal treatment of energy
conversion processes, to the
minimization of errors due to the finite size of the cells of the grid and
to a spatial resolution close to the nominal one. In the cosmological
framework, the basic PPM technique has been modified to include the
gravitational interaction and the expansion of the Universe.

\noindent We present here the simulation of 
a cubic volume of side $75Mpc$ starting
from $z=30$, and applying AMR within a sub-volume 
of side $7.5 Mpc$, centered on a 
$\sim 2 \cdot 10^{14}M_{\odot}$ galaxy cluster.
We re-simulate this volume under different configurations, as reported
in Tab.1. The mass resolution of Dark Matter (DM) particles ranges from
$6.76 \cdot 10^{8} M_{\odot}$ (v256-3 and v256-4) to $4.32 \cdot 10^{10} M_{\odot}$ (v64-3), corresponding to minimum root grid spatial resolutions
 from $292kpc$ to $1.172Mpc$. The maximum spatial
resolution in the region where AMR is applied is 
$\Delta=36kpc$ in all the simulations except for 
the case of v256-4, where $\Delta=18kpc$. 
All runs are non radiative, and furthermore no treatment of reionization background due to AGN and
or massive stars is modeled here.

In all the above simulations, the galaxy cluster is formed through a major merger at $0.8<z<1$, and visual inspection shows that its perturbed dynamical state stays till later
epochs, due to further accretions. Fig.\ref{fig:mass_z} shows the redshift evolution 
of the total mass and of the gas mass inside the virial cluster region, 
measured according to the spherical over-density centered on
the density peak where the galaxy cluster forms.  

Computations described in this work were performed using the 
ENZO code developed by the Laboratory for Computational
 Astrophysics at the University of California in San Diego 
(http://lca.ucsd.edu).

\bigskip

\begin{figure} 
\includegraphics[width=0.49\textwidth,height=0.55\textwidth]{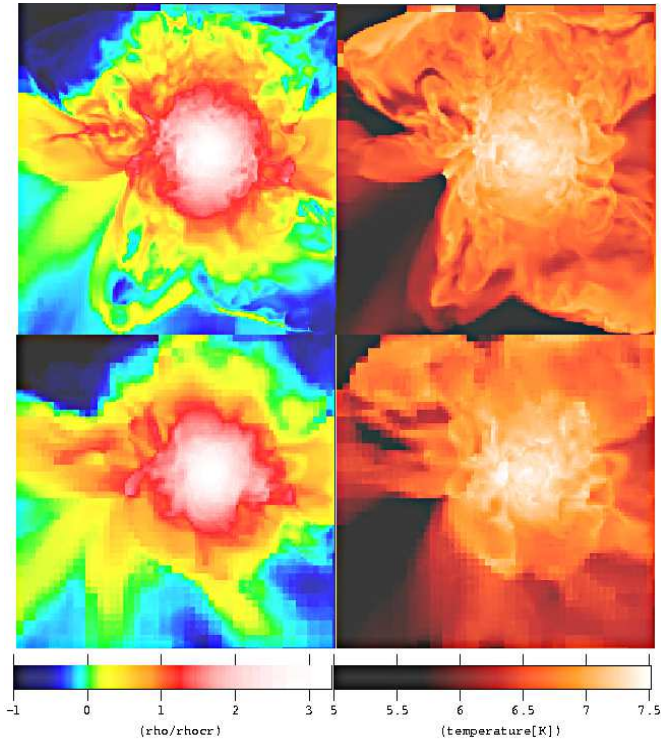}
\caption{Gas density
 and temperature slices for the AMR region of the  
v128-3 run (upper panels), and of the d128 run (lower panel). The gas 
density is normalized to the cosmological critical gas density: $rhocr \equiv \rho_{cr} \equiv \rho_{cr,0} f_{b}$,
where $\rho_{cr,0}=3H(z)^{2}/8\pi G$ is the cosmological critical density 
( $H(z)$ is the Hubble constant),
and $f_{b}$ is the cosmic baryon fraction.
The side of the image is $7.5Mpc$ and the depth along the line of sight is $36kpc$. The gas density is normalized to the 
critical density of the universe, rescaled by the cosmic
baryon fraction.}
\label{fig:amr_maps}
\end{figure}

\section{Adaptive Mesh Refinement technique for Turbulent Motions}
\label{sec:AMR}

The first application of AMR to the study of turbulence in the 
inter stellar medium was reported in 
Kritsuk, Norman \& Padoan (2006).
Iapichino \& Niemeyer (2008) applied a
refinement criterion based on the gas velocity field (analyzing curl
and divergence of velocity), in order to study turbulence in
cosmological ENZO simulations.
Motivated by the above results, here we report on
an exploratory study where a grid refinement
scheme based on the analysis of one dimensional jumps in the
velocity field is introduced in ENZO.
In order to apply this method at full power
to ENZO simulations we combine the implementation
of  the standard grid refinement criterion, customary adopted in cosmological
simulations (based on over density), with a new grid refinement
criterion based on the analysis of the jump of velocity, $\Delta \rm{v}$,
across cells.  This choice ensures that shocks can be studied
with the highest available resolution in simulated galaxy clusters,
contrary to usual AMR runs, and at the
same time this refinement scheme allows to increase the spatial
resolution around turbulent features in the simulated galaxy clusters.

In Teyssier (2002) 1--D tests are presented to assess
the importance of refining shocks (according to
a pressure criterion) in cosmological simulations. The major
findings were that: a) relevant numerical instabilities
do occur when simulated shocks move from a low resolution to a high
resolution region (e.g. from a low density to a high
density environment); b) since most of cosmological shocks move from
high density (collapsing) regions to low density regions,
explicitly refining on shocks can be safely avoided in the run
time calculation of expanding accretion shocks (such as those
developed in a standard Zeldovich pancake collapse).
However, in the high
resolution simulation of galaxy cluster we present here,
we expect significant departures from any idealized self-similar
model of shocks (e.g. Molnar et al.2009), 
and the additional refinement
on shock waves is a interesting option.

In more detail, we propose to use the normalized 1--D velocity jump
across 1--D patches in the simulation, $\delta \equiv |\Delta \rm{v}/v_{m}|$ 
(where $\rm{v}_{m}$ is the minimum velocity, in modulus, over the cells in the patch) 
to trigger
grid refinement.
Even if this method is highly simplified respect to that employed in 
Iapichino \& Niemeyer (2008), in the next Sections we will show that 
it produces a significant step forward, with no significant extra expense
of computational effort, in both the study of shock waves
and the spectral characterization of the gas velocity field inside
galaxy clusters.

In particular we
recursively analyze the velocity jumps
across three adjacent cells at a given AMR level, and
increase the resolution (by a factor 2 in cell size) for the cells of
the patch whenever $\delta$
is larger than a threshold value.
At the same time, also the standard AMR method triggered by gas/DM 
over-density is applied (e.g. Norman et al.2007); the over-density
threshold is set $\delta_{\rho}=\delta \rho/\rho = 2$  (where $\rho$ can
be either Dark Matter of gas density) for all runs. We notice that this 
threshold is smaller
than what usually taken in similar works (e.g. $\delta_{\rho}=4$ in 
O'Shea et al.2004; Nagai et al.2006; Iapichino \& Niemeyer 2008.)
and thus typically much more volume is refined in the simulations 
presented here.

We adopt as reference value $\delta=3$ and allow 
for  a number of AMR levels up
to the maximum resolution of $\Delta=36kpc$.
In one case, we also perform a run using the same setup 
of the v256-3 run, but allowing for 
one more AMR level (4 levels instead of 3), reaching 
the maximum resolution of $18kpc$ (v256-4).
Finally, we present results for $\delta=10$ (v128-10) 
and $\delta=1$ (v128-1), in order
to assess the convergence of our results 
(Sec.\ref{subsec:energy}-\ref{subsec:pk}). 

\noindent A reference simulation is also produced where only
the gas/DM over-density criterion is used to trigger mesh refinements 
(d128), along with a test run where
the AMR criterion triggered by velocity jumps 
is added to the standard one only starting from $z \leq 2$ 
(v128-z2). The latter run is designed in order to establish 
whether it is feasible to apply the novel mesh refinement
criterion starting only at later cosmic epochs, 
where clusters formation starts,
saving some computational effort.

\begin{figure} 
\begin{center}
\includegraphics[width=0.48\textwidth]{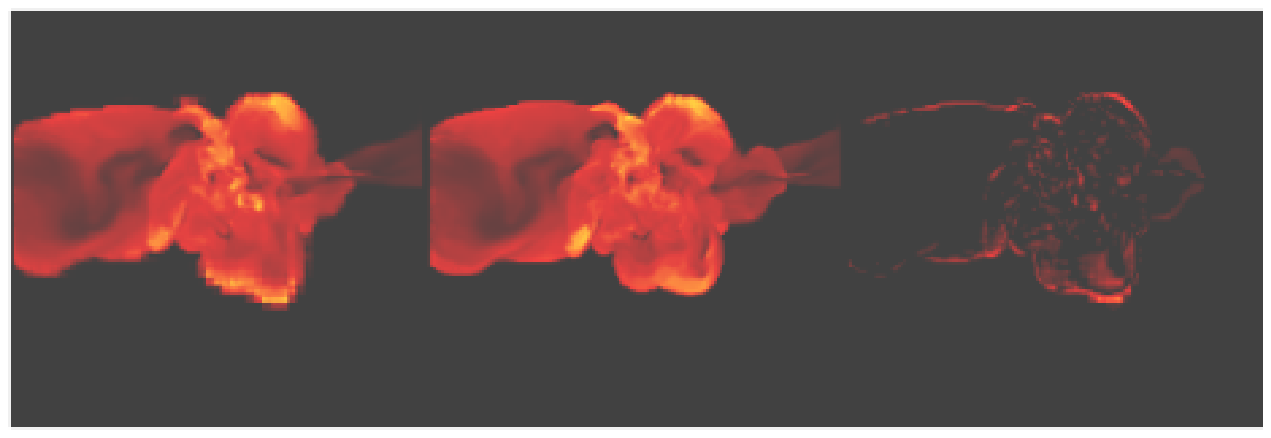}
\includegraphics[width=0.48\textwidth]{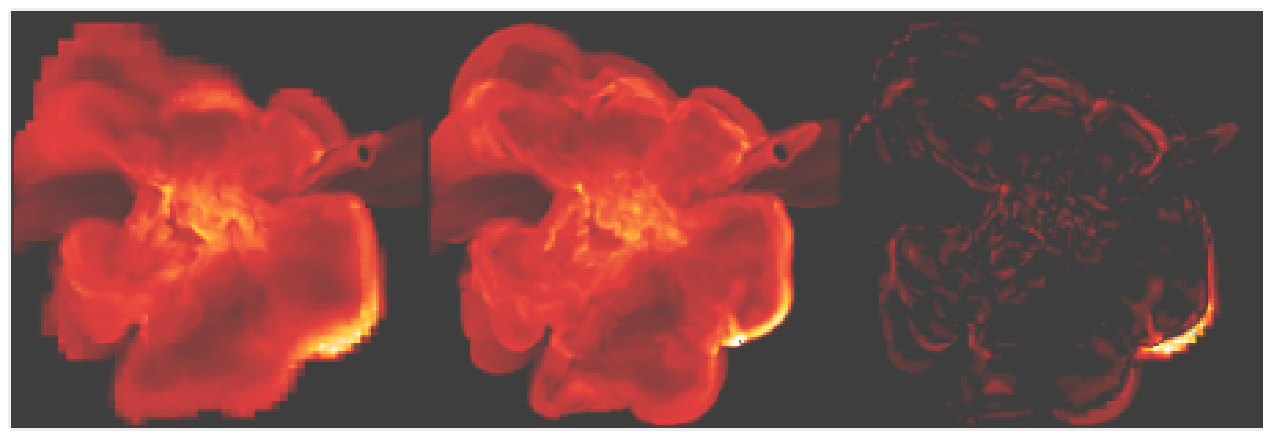}
\includegraphics[width=0.48\textwidth]{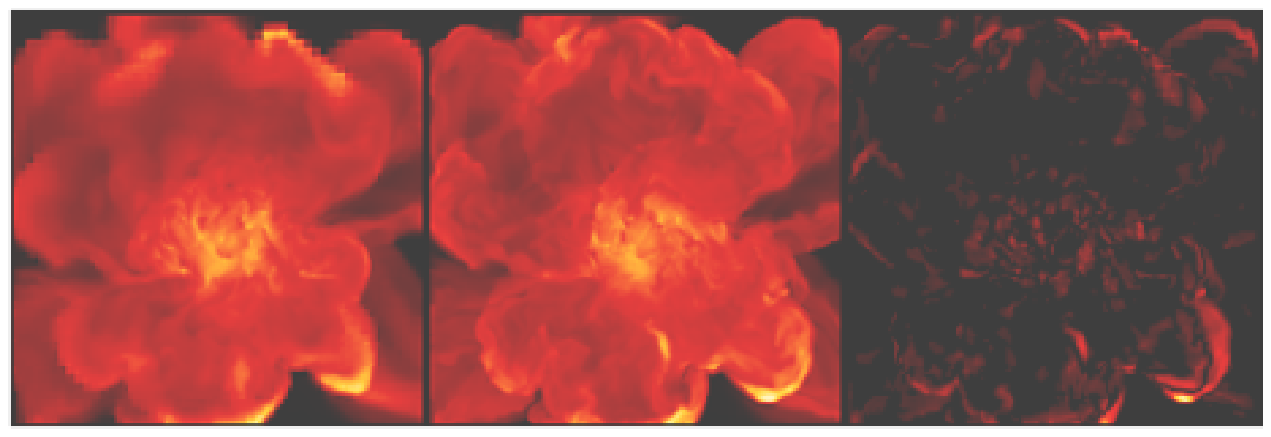}
\caption{Temperature maps for a central slice in the simulated AMR, at 
for different redshifts (z$=2$, z$=1$ and z=$0.2$) by using the 
standard AMR criterion (d128 run, {\it left} panels), the new AMR criterion (v128-3 {\it center}
panels); the {\it right} panels show the cell by cell difference, as $T_{new}-T_{standard}$. The color table is as in Fig.\ref{fig:amr_maps}, {\it right} panels.}
\label{fig:temp_diff}
\end{center}
\end{figure}

\begin{figure} 
\includegraphics[width=0.49\textwidth,height=0.42\textwidth]{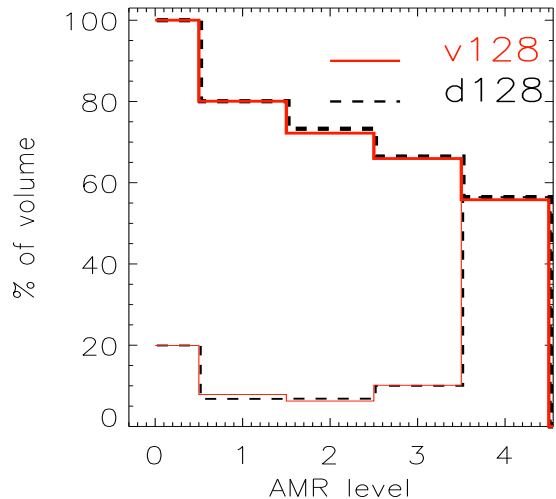}
\caption{Ratio of the volume covered by
cells at the various AMR levels (normalized to the volume of the AMR
region), for runs v128-3 (red-solid) and d128-3 (black-dashed) at $z=0.1$. 
The thick lines gives the cumulative distribution, while the thin lines show the differential distribution. The root grid level is labeled as ``0''.}
\label{fig:amr_levels}
\end{figure}

\section{Results}
\label{sec:results}

\subsection{Comparison with Standard AMR Runs.}
\label{subsec:cfr_amr}

Fig.\ref{fig:amr_maps} shows 2--D slices of gas density and temperature
comparing runs v128-3 and d128 at $z=0.1$.
Unlike in the standard mesh refinement triggered by gas/DM over-density, 
with the new AMR criterion
shocks and chaotic motions are followed at the highest available resolution
in the run, 
$\Delta=36kpc$, up to large ($\sim 3-4 Mpc$) distances from the 
cluster center.
The difference between the two approaches is remarkable at all
stages in the evolution of the cluster, and especially in 
highlighting strong shock waves excited during the major merger event, 
as shown in the temperature
maps of Fig.\ref{fig:temp_diff}. 

In Fig.\ref{fig:amr_levels} we show the distribution of the volume
occupied by cells at the different available AMR levels, comparing the results from  v128-3 and d128 at $z=0.1$.
This shows that the application to the AMR criterion
triggered by velocity jump does not cause any appreciable increase of
memory expense in cosmological numerical simulations, compared to the
adoption of the standard AMR criterion.
Interestingly enough, although the volume occupied by cells at the 
highest AMR level is similar in both runs (i.e. $\sim 55$ per cent of the AMR volume of 
side $7.5$ Mpc), we 
measure a slightly larger number of refined cells in the d128 run ($\sim 1-2$ 
per cent) at all AMR level. Differences in the distributions are consistent 
with the effect driven by the differences in the thermodynamics of the
gas simulated with the two approaches. 
Indeed, 
the v128-3 run has a larger amount of turbulent energy (Sec.\ref{subsec:energy})inside
the simulated galaxy cluster and this reduces
the innermost gas density compared with that of d128;
this balances the larger number of cells triggered according to their velocity jump.

\begin{figure*} 
\begin{center}
\includegraphics[width=0.98\textwidth]{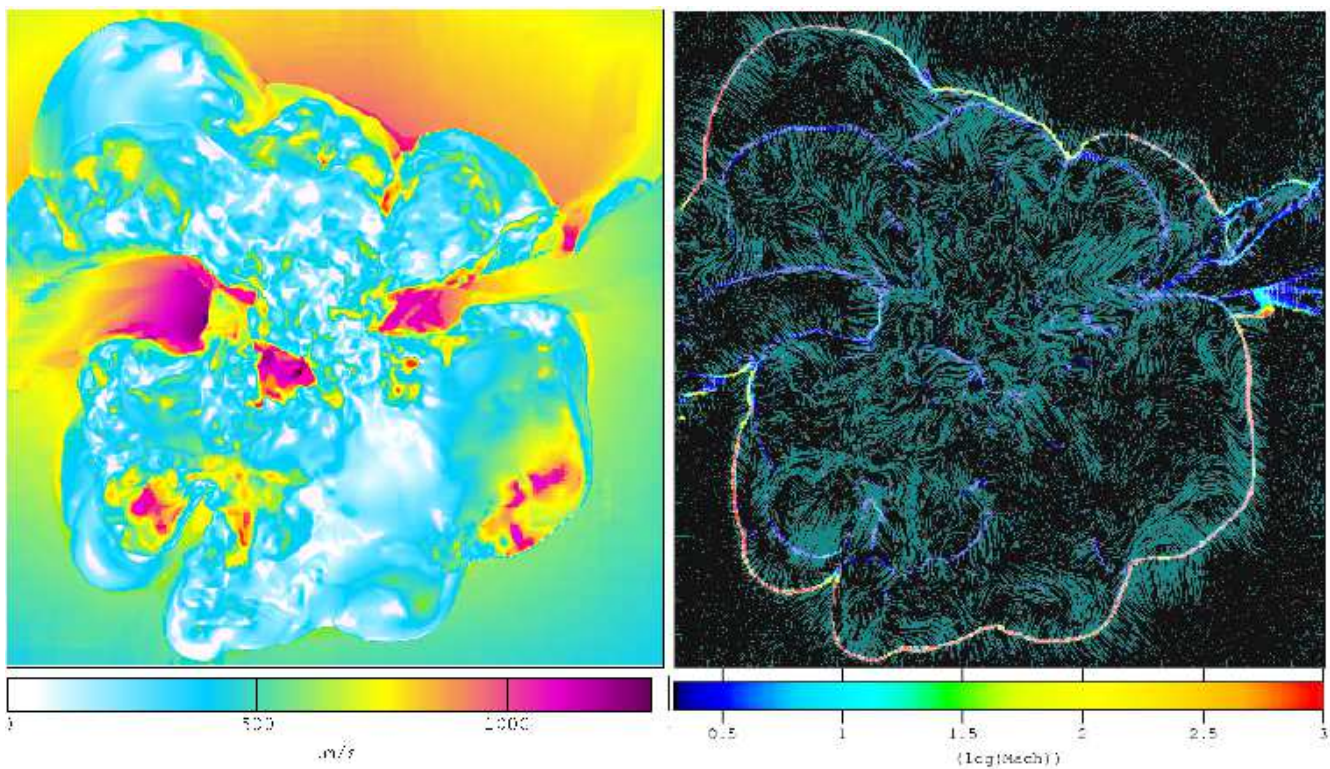}
\caption{{\it Left:} modulus of total gas velocity in a slice of side $7.5Mpc$
and depth $18kpc$, for the v256-4 run at $z=0.6$. {\it Right:} map of Mach
number (in colors) and turbulent gas velocity field (arrows).} 
\label{fig:maps}
\end{center}
\end{figure*}

\begin{figure*}
\begin{center} 
\includegraphics[width=0.99\textwidth,height=0.49\textheight]{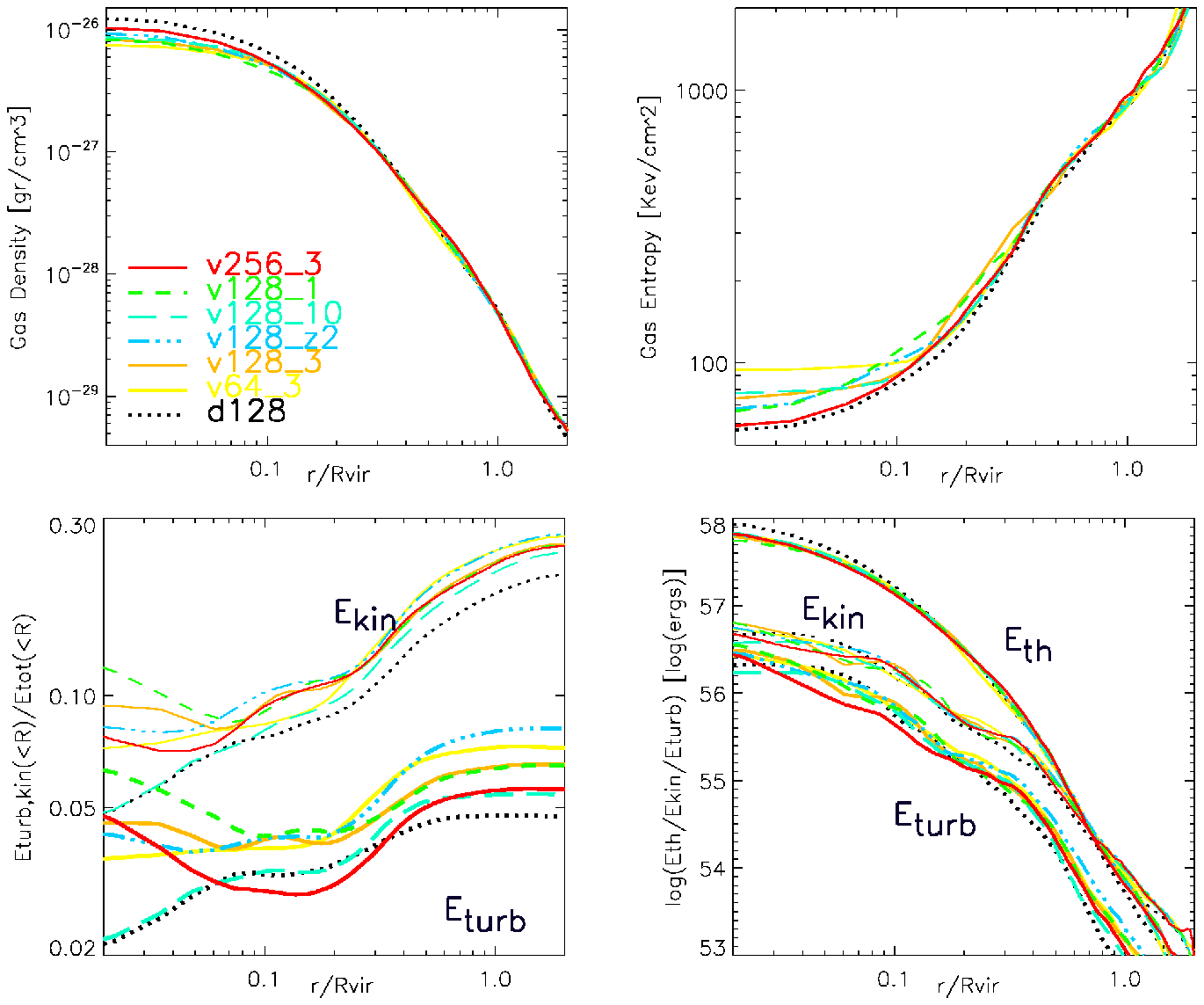}
\caption{Gas density profiles ({\it top left}), gas entropy profiles ({\it top
right}), $E_{turb}/E_{tot}$and $E_{kin}/E_{tot}$ profiles ({\it lower left}) and $E_{turb}$,$E_{kin}$ and $E_{therm}$ profiles ({\it lower right})
for all runs of the paper at $z=0.1$. The color coding is all the same as in the first
panel; $E_{turb}$ refers to gas motions on scales $<300kpc$.}
\label{fig:profile}
\end{center}
\end{figure*}

\subsection{Turbulent motions}
\label{subsec:energy}

In order to characterize turbulent velocity 
fields in the complex environment of galaxy clusters,
it is necessary to extract velocity 
fluctuations from a complex distribution of velocities. 
Dolag et al.(2005) proposed that 
the turbulent gas velocity field can be extracted by removing 
a {\it ''local''} mean velocity field, whose value is obtained 
by interpolating the 3--D gas velocity on large enough scales. 
Using this approach, it was shown that the bulk of laminar infall
motions driven by accreted substructures in Smoothed Particles
Hydrodynamics(SPH) simulations
develops at scales of 
the order of $\geq 100-300 kpc$, which indeed corresponds to
the core radii of matter clumps accreted by massive galaxy clusters. 

Following a similar approach, here we use the ENZO implementation
of the PPM scheme (based on parabolic 
interpolations on cells)  to map the 3--D local mean 
velocity field, $V_{L}$, and for each cell we measure the turbulent 
velocity
as $\Delta \rm{v} = \rm{v} - V_{L}$; $\rm{v}$ 
is the gas velocity at the maximum AMR level, while $V_{L}$ is 
measured at a coarser resolution (for the v256-3 and v256-4 runs
this is $\Delta=292kpc$, while for the other runs we consider the AMR level 
corresponding to this scale). We notice that 
this procedure implies a largest possible scale 
of $\approx 300kpc$ for turbulent motions, and therefore in presence
of significant turbulent motions on larger scales our procedure
would lead to a lower estimate on the total turbulent energy budget.
Yet, the influence of our filtering scale in the final estimate of the 
turbulent energy cannot be larger than a factor $\sim 2$. This simply
comes from the comparison of the kinetic energy and turbulent energy profiles 
reported in the last panel of Fig.\ref{fig:profile}, and it is
consistent with tests previously reported in Dolag et al.(2005) e Vazza
et al.(2006). The visual inspection (e.g. Fig.\ref{fig:maps}) further 
confirms that most of the velocity structure present
in the IGM at scales $>300kpc$ is mostly due to laminar infall motions.


In all runs, the total mass of the cluster at the center of the AMR region at $z=0$ is
$M \sim 2.1 \cdot 10^{14}M_{\odot}$, which corresponds to a virial radius
of $R_{vir}=1.4Mpc$.
Panels in Fig.\ref{fig:maps} show the
total and turbulent velocity fields for an epoch just after
the major merger event, $z=0.6$, for a slice crossing the AMR 
region. 
The laminar infall patterns, due to accretion of sub-clumps from filaments
({\it left} panel),
are almost completely removed by our filtering of the velocity field, and 
small scale curling motions 
injected by accreted clumps and by shocks (see also Sec.\ref{subsec:shocks}) 
are well highlighted ({\it right} panel).

\noindent The uppermost panels in Fig.\ref{fig:profile} show the gas density profile 
and the gas entropy
profiles of the cluster in all runs.
The lower panels in the same Figure show the profiles of thermal, turbulent and 
kinetic energy, and the ratio 
between turbulent (or kinetic) energy and the total energy $E_{tot}$ (kinetic plus thermal) inside a given radius. 
The turbulent energy, $E_{turb}$, is measured as $\rho \Delta \rm{v}^{2}/2$, 
the total kinetic energy is $E_{kin}=\rho \rm{v}^{2}/2$ and 
the thermal energy in the cell is
$E_{th}=(3/2)k_{B}\rho T/\mu m_{p}$;
the velocity field is always corrected for the galaxy cluster center of mass
velocity; all profiles refer to $z=0.1$.
The standard AMR run (i.e. over-density based refinement, d128) shows the 
highest central density and the steepest entropy profile, while
all runs with velocity/over-density refinement have flatter profiles.
This is explained because merger shocks in runs with the velocity/over-density
AMR criterion are simulated with higher accuracy during cluster 
lifetime, and they can propagate
more deeply towards the inner regions of the cluster without
being damped by resolution effect.
At all radii, the runs with the velocity/over-density refinement show larger
energy budget in turbulent motions, with a $E_{turb}/E_{tot} \sim 3-4$ 
percent at 
$r=0.1R_{vir}$ ($E_{turb}/E_{th} \sim 5$ per cent 
within the same radius) and 
$E_{turb}/E_{tot} \sim 5-8$ percent inside $R_{vir}$ ($E_{turb}/E_{th} 
\sim 10-20$ per cent 
within the same radius).
As expected the adoption of a larger threshold for $\delta$ (v128-10)
decreases the budget of turbulent motions in the simulated volume,
gradually approaching the results of standard AMR(d128), except for the
outermost regions, where strong shocks occur and the threshold $\delta=10$
still triggers refinement. 
Decreasing $\delta$ (v128-1) increases the turbulent energy budget, yet 
convergence is already reached at $\geq 0.2R_{vir}$
for $\delta=3$ (v128-3).
The adopting the mesh refinement criterion based on velocity jumps
for $z<2$ (v128-z2) produces profiles consistent with those from
the run where this criterion is applied since the beginning of
the simulation (v128), provided a small difference is found for
$r \sim R_{vir}$.

In the cases where the AMR peak resolution is fixed at $\Delta=36kpc$
(v256-3,v128-3,v64-3),
the adoption of a larger mass resolution in DM particles 
causes a significant decrease in the turbulent budget
at large radii (the kinetic energy profiles,
however, are almost unaffected by that). We find that the reason
for this is that 
in the cluster outskirts, where strong accretion shocks are
located, satellites with a too coarse DM mass resolution have a typically
smaller gas and DM density concentration, and they are more easily stripped
and/or destroyed generating more small scale chaotic motions in the peripheral regions
of clusters (see also Sec.\ref{subsec:shocks}).

The total kinetic energy within $R_{vir}$ in our simulations 
is in line  with SPH results with reduced artificial viscosity (Vazza et al.2006) and 
other AMR results obtained with ENZO (Iapichino
\& Niemeyer 2008). However, it is 
unclear if the observed inner turbulence profile can be
reconciled with with SPH findings, where 
an increase of the ratio between turbulent energy
and the total one is observed with decreasing radius, for $r/R_{vir}<0.1$
(Dolag et al.2005).
On one hand it seems that the progressive increase of the DM mass
and force resolution in our simulations causes the same kind of 
steepening also in our innermost profile, 
on the other hand the turbulent energy budget
remains smaller by a factor $\sim 5-6$ respect to SPH results. 
Whether or not this is related 
to the different clusters 
under observation (and to their 
dynamical states) or if this is this a more fundamental issue caused
by differences between AMR and SPH simulations, is a topic that deserves more
accurate investigations in the future.

\begin{figure*}
\begin{center} 
\includegraphics[width=0.49\textwidth,height=0.49\textwidth]{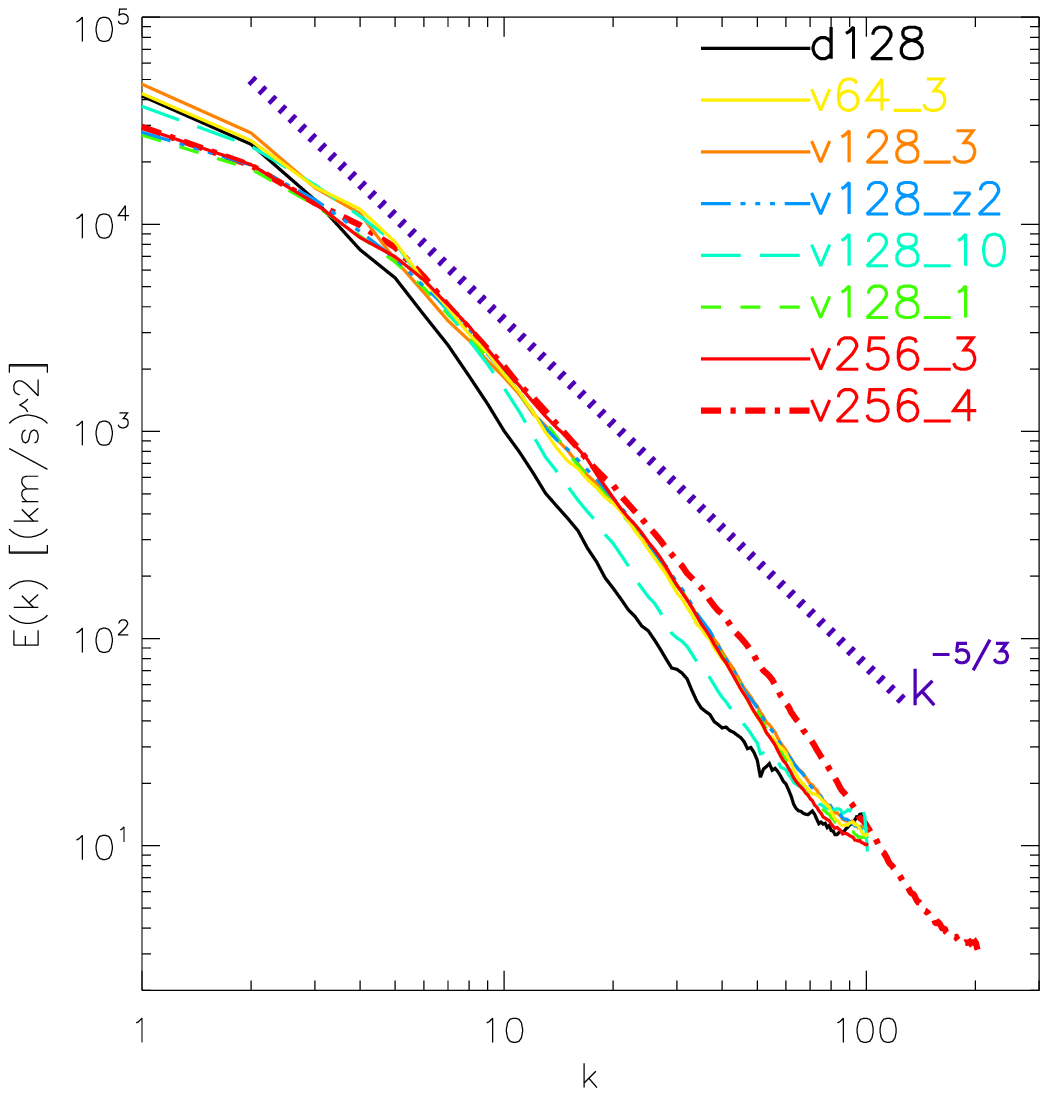}
\includegraphics[width=0.49\textwidth,height=0.49\textwidth]{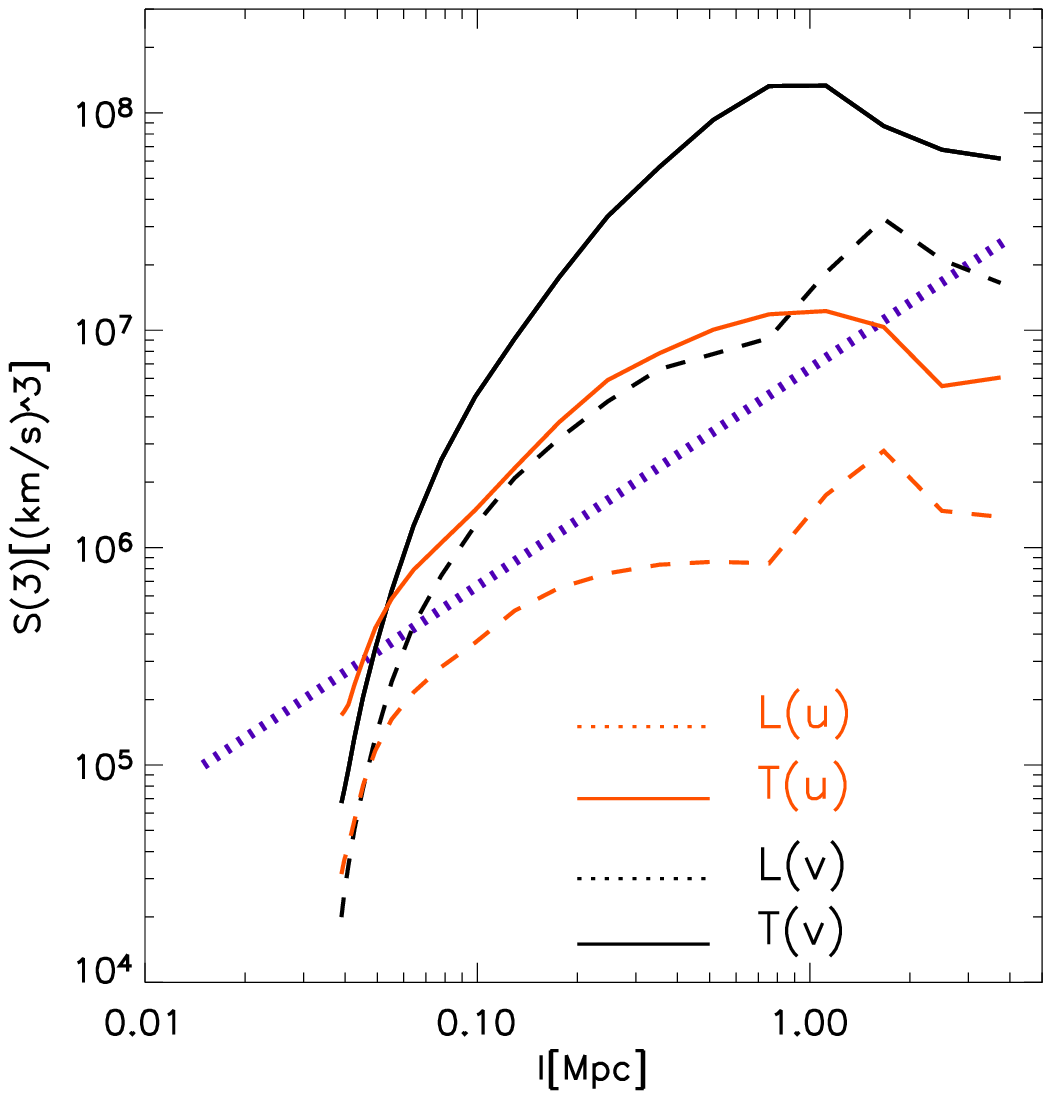}
\caption{{\it Left:} 3D power spectra for the velocity field of the various run at $z=0.1$. 
The spectra are shown up to their Nyquist frequency; the purple dashed
lines shows the $-5/3$ slope to guide the eye.
 {\it Right:} longitudinal and transverse third--order structure functions for velocity field, $\rm{v}$ ({\it black}, and for the density=weighted velocity field, $u \equiv \rho^{1/3} \rm{v}$ ({\it red}), for a sub volume in the v128-3 run. The purple dashed line shows the expected scaling for the Kolmogorov model.}
\label{fig:power}
\end{center} 
\end{figure*}

\subsection{Power Spectra and Structure Functions of the Turbulent Velocity Field}
\label{subsec:pk}

We characterize the cluster velocity field through it 3D power
spectrum, $E(k)$, defined as:

\begin{equation}
E({\bf k}) = {1\over 2} |{\bf \tilde{v}(k)}|^{2},
\end{equation}

where ${\bf{\tilde{v}(k)}}$ is the Fourier transform of the 
velocity field:

\begin{equation}
{\bf {\tilde{v}(k)}}=
\frac{1}{(2\pi)^3}\int_{V}{\bf {v}(x)}e^{-2\pi i \,{\bf{k \cdot x}}}d^{3}x.
\end{equation}


$E(k)$ is calculated with 
standard FFT algorithm (e.g. Federrath, Klessen \& Schmidt 2009, and
references therein),
and with a zero-padding technique to deal with the non-periodicity
of the considered volume .
Differently from SPH
and standard AMR simulations, the velocity plus density refinement 
allows to follow the cluster velocity field 
with high spatial resolution
in lower density regions, with  important consequences on the capability to
describe its spectral properties over a wide range of scales.

The {\it left} panel in Fig.\ref{fig:power} shows the 3--D power spectra calculated for all runs at $z=0.1$.
$E(k)$ is approximately described by a simple power law over
more than one order of magnitude in $k$, with a slope not
far from a standard Kolmogorov model ($E(k) \propto k^{-5/3}$). 
At large scales ($k<4$) a 
flattening in the spectrum is observed in all runs, 
at a wave number roughly corresponding to the virial 
diameter of the cluster, 
which likely
identifies the outer scale of turbulent motions connected with
accretion processes.
We remark that for spatial scales $\leq 32 \Delta$, the
slope of the power spectrum is affected
by the non-uniform numerical dissipation that 
PPM adopts to increase resolution
in shocks and contact discontinuities (Porter \& Woodward, 1994).

\noindent As in the case of the turbulent energy budget, the v128-10 run
falls in between the standard AMR run and all the other runs with
velocity/over-density refinement, while there is almost no difference by adopting $\delta =3$ or $\delta = 1$ as threshold; we find no relevant 
differences if the velocity refinement criterion is adopted at $z<2$ (v128-z2)
or at $z=30$ (v128).

Remarkably due to its larger peak resolution,
the v256-4 shows a regular power law for almost two orders
of magnitude, thus supporting the picture that
the simulated IGM is globally turbulent starting from
sub--Mpc scales. This is also further suggested by the 
{\it right} panel in Fig.\ref{fig:power}, which shows the third order 
velocity structure functions 
for the v128-3 run, calculated as 

\begin{equation}
S_{\it p}(\it l)=<|{\bf v(r+l)-v(r)}|^{3}>.
\end{equation}

Shown are the transverse (${\bf v \perp l}$) and longitudinal (${\bf v \parallel l}$) structure
functions extracted from a random sub-sample of $\sim 10^{5}$ cells 
in the simulated volume. For completeness, structure functions are also calculated 
for the density-weighted velocity, ${\bf u} \equiv \rho^{1/3} {\bf v}$, 
which was introduced by Kritsuk et al.(2007) to study scaling relations
for simulated supersonic turbulence.
All signals show a peak at $\sim Mpc$ scales, 
thus implying that the maximum outer scale 
that drives turbulence is of the order of $R_{vir}$.


\begin{figure} 
\includegraphics[width=0.49\textwidth]{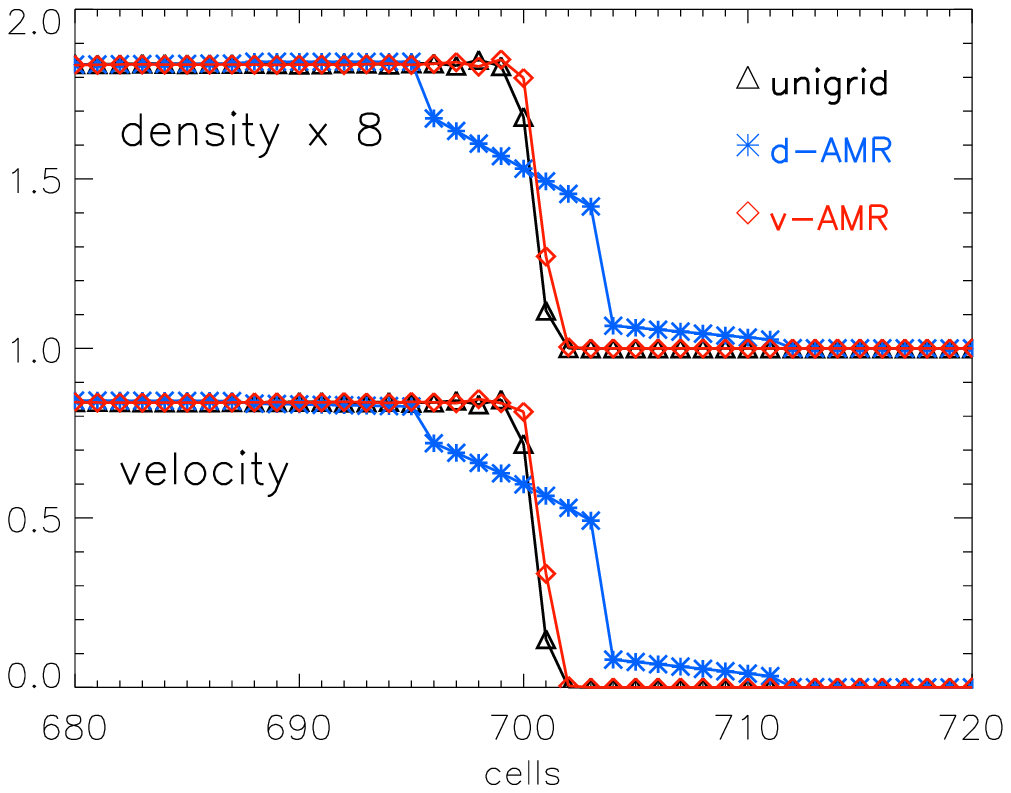}
\includegraphics[width=0.49\textwidth]{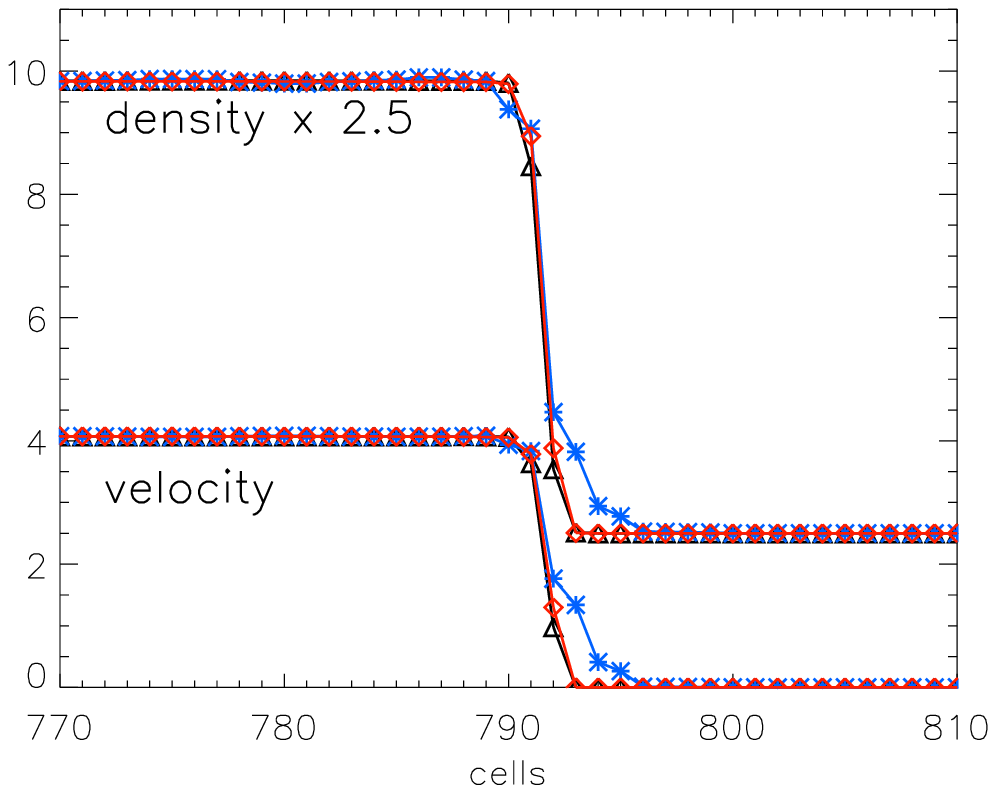}
\caption{{\it Top panel:} gas density and gas velocity for a 1-D shock tube
test, at the position of a $M \approx 1.5$ shock wave. {\it Bottom panel:}
gas density and gas velocity for a 1-D shock tube test, at the
position of a $M>>10$ shock wave. 
In both cases results are
shown for a run with $N=1024$ cells with uniform
resolution ({\it unigrid}), and for two AMR runs
with root grid $N=128$ and 3 additional level of
refinement, triggered by density jump ({\it d-AMR})
and by velocity jump ({\it v-AMR}).
The values of gas densities in both cases have been rescaled
by an arbitrary value to avoid overlapping with the other
lines. }
\label{fig:tube}
\end{figure}

\begin{figure} 
\includegraphics[width=0.45\textwidth]{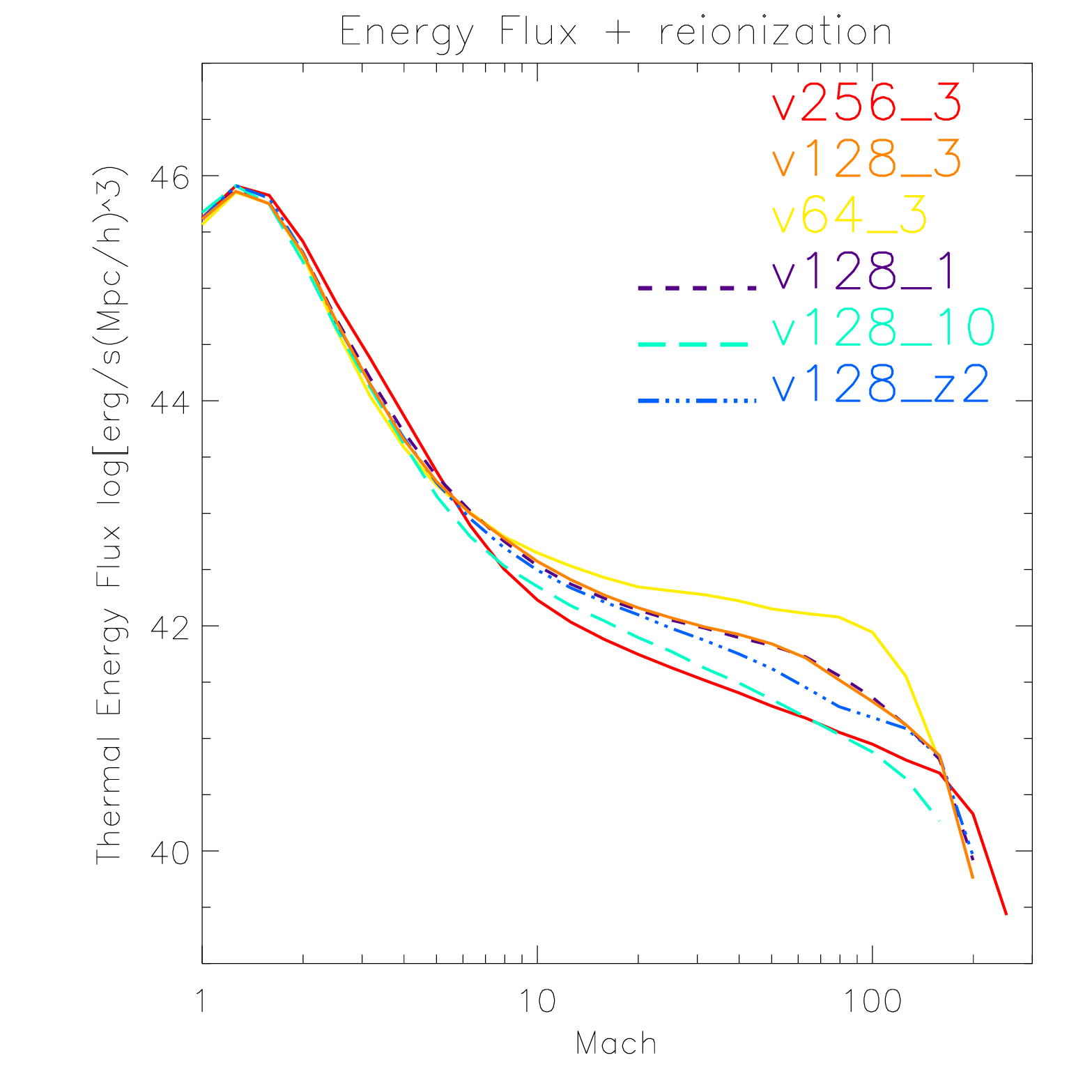}
\includegraphics[width=0.45\textwidth]{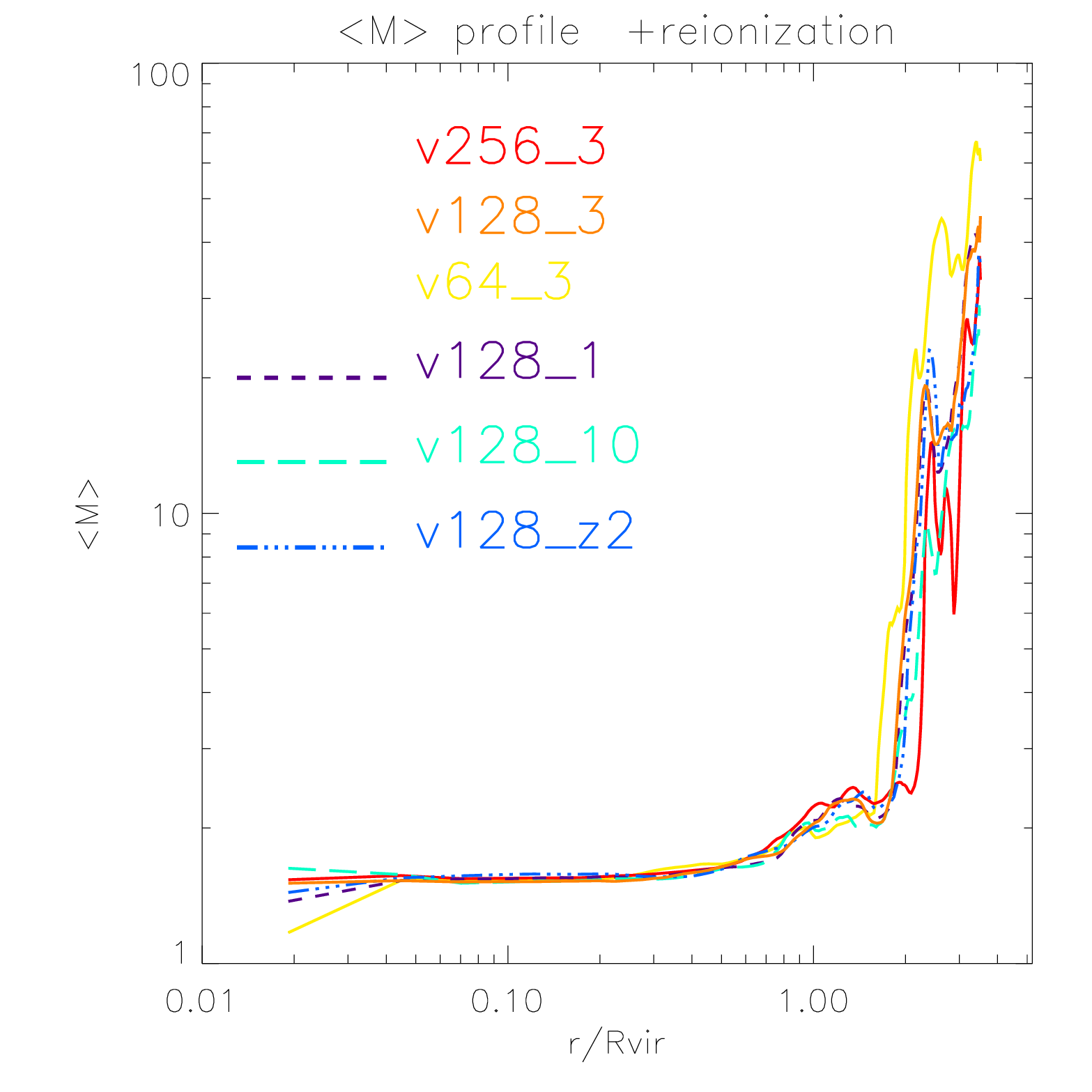}
\caption{{\it Top:} distribution of thermal
energy flux at shocks within the AMR region of all simulated runs with the
over density/velocity AMR criterion, at the grid resolution of $\Delta=36$ kpc.
{\it Bottom:} radial profile of the 
energy-flux weighted average Mach number at z=0.1, 
for all runs with the over density/velocity AMR criterion.}
\label{fig:hist_th}
\end{figure}

\subsection{Shock Properties}
\label{subsec:shocks}

Shocks in large scale structures have 
been investigated in a number
of semi-analytical (Gabici \& Blasi 2003; Berrington \& Dermer 2003; 
Keshet et al.2003) and numerical works (Miniati et al.2001; Ryu et al
2003; Pfrommer et al.2007; Skillman et al.2008, Vazza et al.2008; Molnar
et al.2009).  
Observationally, merger shocks have been detected only in a few 
in a few nearby X-ray bright galaxy clusters (Markevitch et 
al.2005; Markevitch 2006; Solovyeva et al.2008), and 
may be possibly associated with single or double radio relics discovered
in a number of galaxy clusters (e.g. Roettgering et al.1997; Markevitch et al.2005; Bagchi et al.2006; Bonafede et al.2009; 
Giacintucci et al.2008).

The application of the AMR approach described in this paper to galaxy
allows to follow
with high resolution the onset and the evolution of shock waves
in the IGM of simulated galaxy cluster within $\sim 2 R_{vir}$ 
from the clusters center, and to explore the connection 
between shocks and turbulence in large scale
structures.

In Fig.\ref{fig:tube} we present standard 1--D shock-tube
tests for a weak ($M \approx 1.5$) and for a strong ($M >> 10$) shock, 
where we compare the application of the AMR
criterion based on velocity jumps (with $\delta=3$), the
application of the density jump criterion (with $\delta_{\rho}=2$),
and a simulation with constant spatial resolution fixed
at the maximum resolution level of the AMR runs.
In both tests, the run with AMR based on velocity jumps well matches
the results of the fixed resolution run at the position of the
traveling shock waves, and as result the correct shock jump conditions
can be basically measured across 3 cells, in both 
cases (e.g. Tasker et al.2009). 
On the other hand the AMR based on over density 
smears the weak shocks across a larger distance, since the small
density jumps associated with $M \approx 1.5$ is not large
enough to trigger any mesh refinement; however in the case
of the $M>>10$ shock the gas compression
is large enough to trigger two level of refinement also
in the AMR method based on over density.

Shocks in the our cosmological simulations are identified
by means of the procedure presented in Vazza, Brunetti \& Gheller (2009).
The algorithm works in the following steps:

\begin{itemize}
\item we consider candidate shocked cells those with 
$\nabla \cdot {\rm v} < 0$ (calculated as 3--dimensional velocity 
divergence to avoid confusion
with spurious 1--dimensional compressions that may happen
in very rarefied environments);

\item since shocks in the simulation are typically spread over
a few cells, we define the shock center 
with the position of the
cell in the shocked region with the minimum divergence;

\item we scan the three Cartesian axes with a one--dimensional procedure 
measuring the velocity jump, $\Delta \rm{v}_{x,y,z}$, between 3
cells across the shock center;

\item the Mach number of the shock is obtained by inverting

\begin{equation}
\Delta \rm{v} =\frac{3}{4}c_{s}\frac{1-\it{M}^{2}}{\it{M}^{2}},
\end{equation}

where $c_{s}$ is the sound speed
in the pre-shock region (the cell with the minimum temperature);

\item we finally reconstruct the 3-D Mach number in shocked cells as 
$M = (M_{x}^{2}+M_{y}^{2}+M_{z}^{2})^{1/2}$, that would  minimizes 
projection effects in the case of diagonal shocks{\footnote{We notice that recently Skillman et al.(2008) pointed out that
the measure of Mach numbers with a temperature-based method using a coordinate
split algorithm, overproduces the number of shocks in the case of complex,
oblique flows.}}.

\end{itemize}

In order to have a realistic value of the shock Mach number at accretion
shocks, we apply the post-processing re-ionization scheme in 
Vazza, Brunetti \& Gheller (2009), by increasing the
gas temperature inside cells according to:

\begin{equation}
T_{min}(K)= 450 \,\,( {{\rho}\over{\rho_{0}}})^{0.60},
\label{eq:ion}
\end{equation}

(where $\rho$ is the gas density and $\rho_{0}=10^{-32}$gr cm$^{-3}$)
which is found to mimic the integrated effect of the 
Haardt \& Madau (1996) run-time re-ionization scheme adopted in ENZO with
sufficient accuracy.

The Right panel in Fig.\ref{fig:maps} shows a map of reconstructed
Mach number with this method, for a slice of $18kpc$ crossing the 
AMR region of run v256-4 (with overlaid streamlines of the turbulent
velocity field).

In the case that AMR is forced to increase the
spatial resolution also around shocks, as our simulations 
with the novel AMR criterion, the Vazza et al.(2009) shock detecting scheme
can be straightforwardly applied to simulated data at the highest 
available AMR level (therefore excluding our d128 run). We thus 
analyze shock
statistics in all runs employing the over density/velocity 
AMR criterion at the resolution of $\Delta=36kpc$.

The thermal energy flux across shocks is customary evaluated as:

\begin{equation}
f_{th} = \delta_{M}(M) \cdot \rho M^{3} v^{3}_{s}/2,
\end{equation}

where $\rho$ is the pre-shock density and $\delta_{M}(M)$ is a 
monotonically increasing function of $M$ (e.g. Ryu et al.2003).

Fig.\ref{fig:hist_th} ({\it top} panel) shows the distribution of thermal
energy flux at shocks within the AMR region. The distribution
of weaker (mostly internal) shocks peaks at $M \sim 1.5$. Overall, 
the distributions are very steep and consistent with those reported in
Vazza, Brunetti \& Gheller (2009). Compared to Pfrommer
et al.(2007), who studied shock energetics with high resolution GADGET2
simulations, we find significantly steeper energy flux distributions in
all our runs,
 $\alpha_{th}\approx -3.5$ (with $f_{th}(M)M \propto M^{\alpha_{th}}$)
for $M<10$, compared to $\alpha_{th}\approx -2$ within the same range
of Mach number in Pfrommer et al.(2007). 
The distributions of the various re-simulations show 
relevant differences only for
shocks with $M>10$, where two clear trends can be found:

\begin{itemize}

\item for a fixed DM mass resolution, the adoption of $\delta=10$ (v128-10)leads to a 
significant reduction of the thermal energy flux at strong shocks compared to
the other runs;

\item for a given AMR criterion based
on over density/velocity jumps, increasing the DM mass
resolution leads to a significant decrease in the thermal energy flux
processed at strong shocks and to a progressive steepening
of the thermal energy flux distribution.

\end{itemize}

Overall we conclude that, once that the velocity jump AMR criterion
is adopted, the largest amount of difference in the spatial
and energy distribution of shocks is caused by the DM mass
resolution.

\begin{figure*} 
\begin{center}
\includegraphics[width=0.9\textwidth]{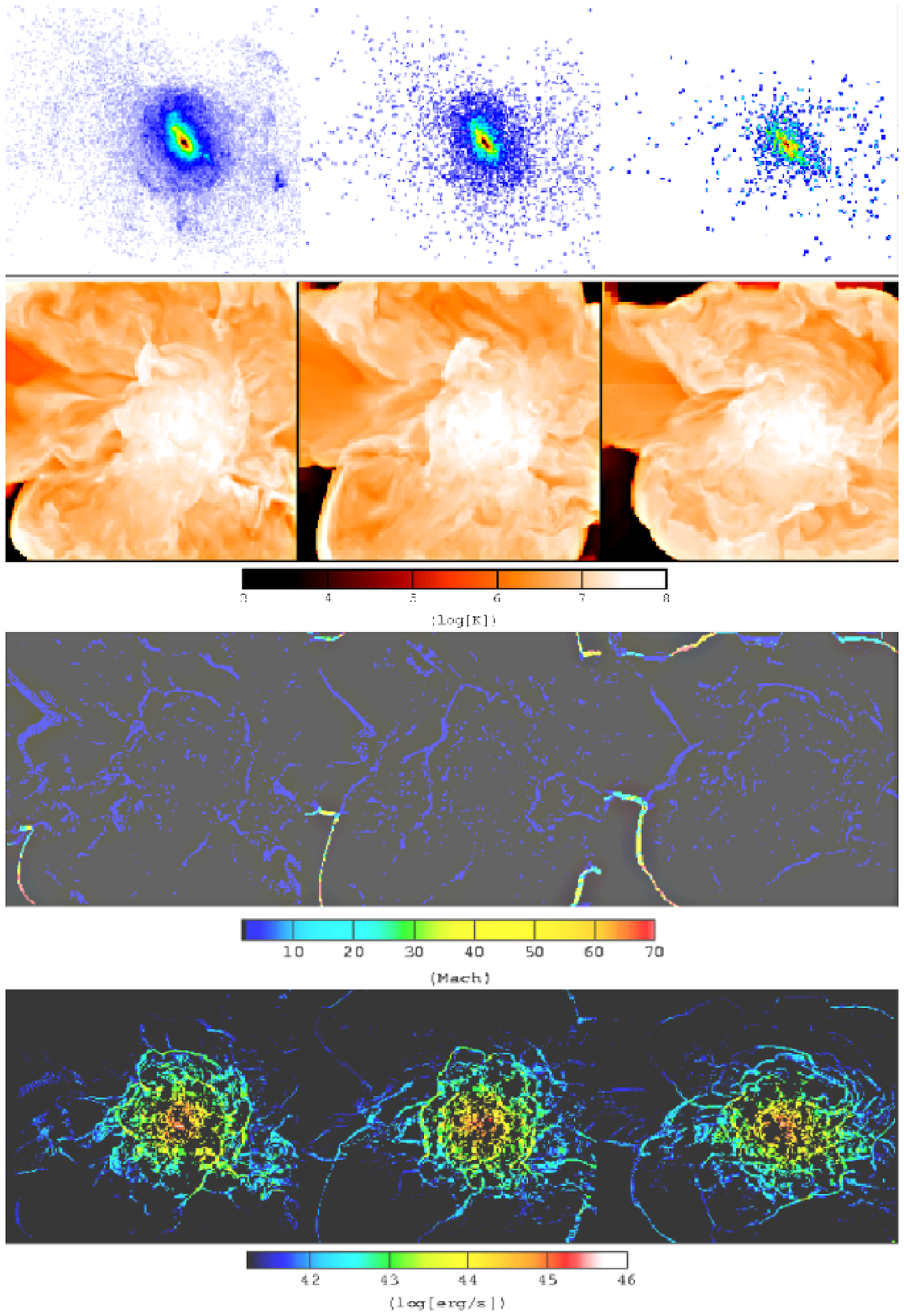}
\caption{From top to bottom: a) projected Dark Matter density maps; b) 
projected volume weighted temperature maps; 
c) projected thermal energy flux-weighted
maps of Mach number; d) projected maps of thermal energy flux at shocks.
The left column refers to run d256-3, the center one to d128-3 and the
right one to v64-3. Each image has side 7Mpc and a LOS depth of 100kpc.}
\label{fig:maps_fth}
\end{center}
\end{figure*}

In order to highlight the reason for this finding, we show in 
Fig.\ref{fig:maps_fth} the maps of projected Dark Matter density,
gas temperature, mean Mach number and thermal energy flux for a 
slice of depth=100 kpc for runs v256-3, v128-3
and v64-3. 
The increase in the number of accreted DM
clumps in simulations with higher DM mass resolution
is found to 
generate a more complex 
temperature distribution, which follows the pattern of matter
infall on the cluster. On the other hand when the DM mass resolution
is coarse, outer shocks are found to be more regular in shape, and they
are characterized by sharper jumps. 
The decrease of DM mass resolution implies that the cluster
becomes more spherically symmetric due to the lack of substructure, 
thus our findings qualitatively support
those of Molnar et al.(2009), which shows that the importance 
of pressure jumps due to accretion shocks in simulated clusters 
is reduced by a factor 5-10 compared to predictions based
on spherical models (Kocsis et al.2005).

Fig.\ref{fig:hist_th} ({\it bottom} panel) shows the radial profile of the
energy-flux weighted average Mach number for all runs with the
over density/velocity refinement. 
All runs produce consistent profiles up to $R_{vir}$, with $<M> \approx 1.5$.
As seen above, differences are larger at accretion shocks 
outside $R_{vir}$, and
in particular we find that as soon as the 
DM resolution is increased, the mean
strength of shocks at $r \sim 1-2 R_{vir}$ is reduced by a factor 
$\sim 2-5$.

\subsection{Time Evolution}
\label{subsec:ev}

\begin{figure*} 
\begin{center}
\includegraphics[width=0.43\textwidth]{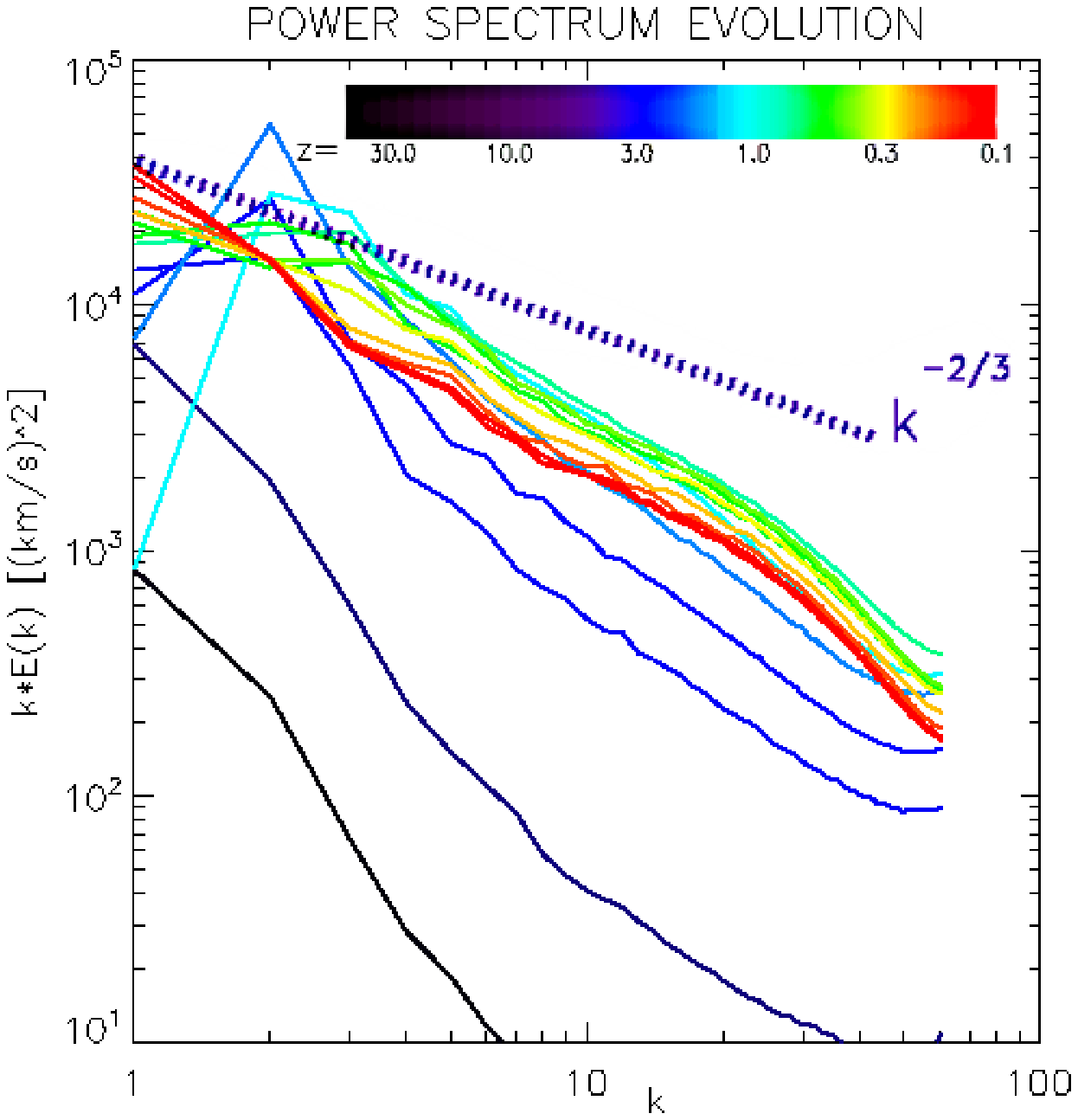}
\includegraphics[width=0.43\textwidth]{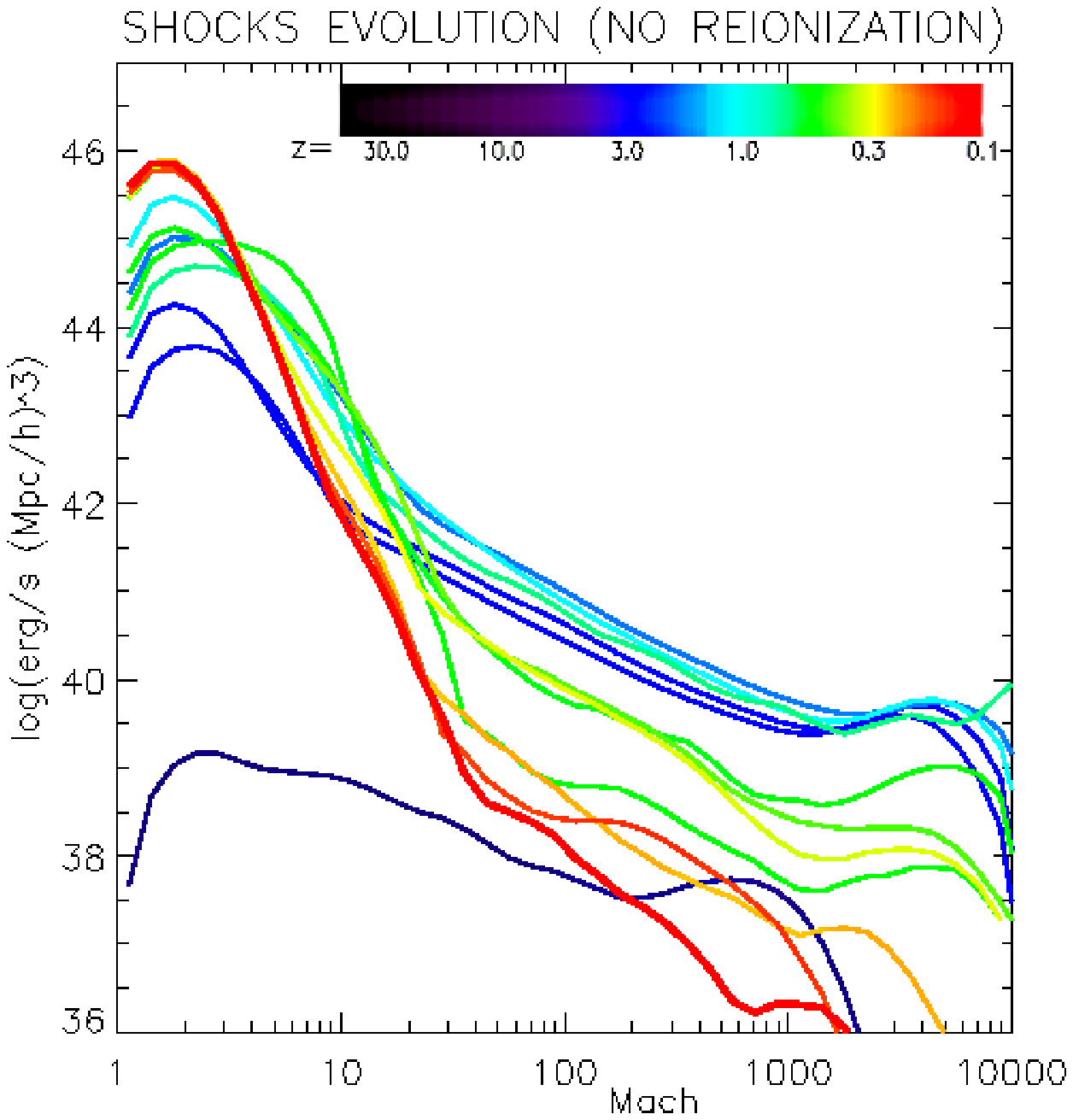}
\caption{{\it Left:} time evolution from of the $k E(k)$ for a sub-volume of side
$3.5Mpc$ in the v256-3 run. The additional dashed line shows the slopes for the Kolmogorov model. {\it Right:} time evolution of the thermal energy flux at
shocks for the same volume. The color coding for the liens is shown in the
color bar.}
\label{fig:pk}
\end{center}
\end{figure*}

We produced a highly time--resolved study of turbulence and 
shocks developing in the AMR region of run v256-3.  
Left panel in Fig.\ref{fig:pk} shows the 
evolution with cosmic time of $k E(k)$ within a sub-volume of $3.5Mpc$ centered on
the cluster center.
The bulk of 
turbulence injection starts with the onset of the major
merger, at $z \sim 1$, and develops at scales in the range $\sim 1-2Mpc$.
At smaller redshifts, the spectrum
gradually approaches the shape reported in Fig.\ref{fig:power}.

To better explore the connection between shock waves and turbulence
in the merger event, we show in Fig.\ref{fig:pk} ({\it right} panel)
the evolution of the thermal energy
flux through shocks for the same sub-volume considered in the Left panel.
The energy flux is calculated with the same procedure as in Sec.\ref{subsec:shocks},
but in this case no treatment of re-ionization is considered, and therefore
accretion shocks are stronger than those measured in 
Fig.\ref{fig:hist_th}, because of the unrealistically
low value of gas temperature outside the galaxy cluster at evolved
redshifts. 

A bump of thermal energy flux at strong merger shocks is measured
at approximately at the same epoch when
the bulk of large scale kinetic energy is injected in the IGM. 
Soon after virialization occurs, extremely strong shocks become rarer and the shocks
energy distribution approaches the distribution of Left panel in 
Fig.\ref{fig:maps_fth} (provided that the considered volume is smaller, and that re-ionization is not modeled here).

\section{Conclusions}
\label{sec:conclusions}

  A very simple implementation of a new refinement criterion in 
ENZO simulations allows
to follow shocks and turbulent motions 
with unprecedented detail up to large distances from
cluster centers.
This refinement criterion is 
successful in catching
the bulk of turbulent motions developed in the IGM
by cluster formation processes, allows 
us to measure
velocity power spectra across two orders of magnitude in spatial scales,
and to follow shocks statistics and  evolution over time in great detail.

Compared to the standard grid refinement criterion, we find that
the extra refinement on velocity jumps causes no significant extra expense
of memory storage, and that by construction it readily allows to use 
accurate shock
detecting scheme at the largest available resolution in these simulations.
In all the analyzed runs, the simulated IGM is found to
host turbulent motions (on scales $<300kpc$) accounting for 
a $\sim 5-25$ per cent
of the gas thermal energy within $R_{vir}$.
Compared to refinement based on over-density only, the new criterion shows 
lower inner gas density, flatter entropy profiles, significantly 
larger turbulence budget at all radii and a larger thermal
energy budget processed at accretion shocks. 
This is due to the sharper representation of shock waves and turbulent motions,
and highlights the importance of highly resolving these phenomena in discussing
accretion processes in the IGM of galaxy clusters.

When the new over density/velocity AMR criterion is employed, the DM mass
resolution is found to play a fundamental role in setting the 
properties of the
turbulence generation and of thermal energy flux at shocks;
if
DM resolution is increased, in-falling matter clumps are less easily
destroyed  during accretion and they thus inject less turbulence via
the ram pressure stripping mechanism. In addition, the complex accretion
pattern established in simulations with high DM mass resolution is found
to significantly prevent the
formation of sharp accretion shocks, compared to runs 
where the DM mass resolution is coarser.
In our simulations we find no relevant differences 
in the properties of turbulence and shock waves if the 
extra refinement based on velocity jumps is considered 
only starting from $z<2$.

Overall, the above results confirm that
shocks, turbulence and dark matter clustering are inter-playing
key ingredients which modern cosmological numerical simulations need to 
follow with high order accuracy and
high resolution to model 
the thermal (and non thermal) properties of the IGM
in a realistic way.

 \section*{acknowledgments}
  F. V. thanks D.Collins, S. Skory and J. Bordner for the support he received while visiting CASS (San Diego), and acknowledges G. Tormen of useful discussions. F. V. thanks M. Nanni and F. Tinarelli for valuable technical support at Radio Astronomy Institute (Bologna). We thanks the anonymous referee for comments which helped us to improve the quality of the paper. We acknowledge partial support through grant ASI-INAF I/088/06/0, and the usage of computational resources under the CINECA-INAF agreement.

\end{document}